\documentclass[twocolumn,tighten]{aastex62}
\usepackage{color}
\usepackage{tikz}

\definecolor{Dblue}{rgb}{0.032,0.032,0.25}
\definecolor{Dgreen}{rgb}{0.032,0.25,0.032}
\definecolor{Dred}{rgb}{0.25,0.032,0.032}
\definecolor{Dora}{rgb}{0.57,0.48,0.124}

\newcounter{note}

\received{xxx, 2020}
\revised{yyy}
\accepted{zzz}

\shorttitle{gravitational noise}
\shortauthors{Larchenkova et al.}

\begin{document}

\title{Influence of the Galactic gravitational field on the positional accuracy
of extragalactic sources.\\ II Observational appearances and detectability}

\correspondingauthor{Tatiana~Larchenkova}
\email{ltanya@asc.rssi.ru}

\author{Tatiana~I.~Larchenkova}
\affiliation{P.N.Lebedev Physical Institute, Leninskiy prospect 53, Moscow 119991, Russia}

\author{Natalia~S.~Lyskova}
\affiliation{National Research University Higher School of Economics, Myasnitskaya str. 20, Moscow 101000, Russia}
\affiliation{Space Research Institute (IKI), Russian Academy of Sciences, Profsoyuznaya 84/32, 117997 Moscow, Russia}
\affiliation{P.N.Lebedev Physical Institute, Leninskiy prospect 53, Moscow 119991, Russia}

\author{Leonid~Petrov}
\affiliation{NASA GSFC, 8800 Greenbelt Rd, Greenbelt, MD 20771 USA}

\author{Alexander~A.~Lutovinov}
\affiliation{Space Research Institute (IKI), Russian Academy of Sciences, Profsoyuznaya 84/32, 117997 Moscow, Russia}
\affiliation{National Research University Higher School of Economics, Myasnitskaya str. 20, Moscow 101000, Russia}

\begin{abstract}

We consider a possibility of detecting the jitter effect of apparent celestial positions of distant sources
due to local fluctuations of the Galaxy gravitational field.
It is proposed to observe two samples of extragalactic sources (target and control) in different sky directions
using the high-precision radio interferometry.
It is shown that on a scale of $\sim$2 years, it is possible to detect a systematic increase
in the standard deviation of measured arc lengths of pairs of target sources
compared to the control ones at the $3\sigma$-level if the accuracy of
differential astrometric observations is around 10 $\mu$as. For the current state-of-the-art accuracy of 30~$\mu$as
achieved at the KVN or VERA interferometers, which have shorter baselines in comparison with VLBI,
the target and control samples will differ only at the 2$\sigma$-level on the scale of 10 years.
To achieve the $3\sigma$-level on this time interval, it is necessary to improve the accuracy up to
$\sim 20~\mu$as.
Other possible effects that can also affect the arc length measurements between two sources are discussed,
and an observational strategy to minimize them is suggested.

\end{abstract}


\section{Introduction} \label{sec:intro}

Before reaching an observer, an electromagnetic radiation of
extragalactic sources propagates through the gravitational field of our
Galaxy. While stationary on large scales, the gravitational field of the
Galaxy is subject to local fluctuations due to motions of stars, compact
relativistic objects and invisible compact halo objects. These
inhomogeneities could lead to different observational appearances, in
particular, to variations with time in an apparent position of any
extragalactic sources (a so-called ``jitter''). In particular, active
galactic nuclei (AGNs) have ultra-compact cores associated with
supermassive black holes (SMBHs). Nowadays, positions of such cores can be
determined with sub-nanoradian accuracies using a very long baseline
interferometry (VLBI, \citealt{r:vcs1}) or space astrometry, e.g. {\it Gaia}
\citep{r:gaia_dr2}. A transversal motion of SMBHs located at
cosmological distances is supposed to be very small, well below
$1~\mu$as/year. The gravitational deflection in the inhomogeneous
non-stationary gravitational field of the Galaxy will cause the jitter in
apparent positions of an AGN core. This effect imposes a fundamental
constraint on the achievable accuracy of high-precision astrometric
observations. Due to the high importance, this topic has been actively
investigated since the nineties of the last century \citep[see, e.g.][and
references therein]{1995A&A...299..321Z,2000ApJ...534..213D,2012ApJ...757..189Y, 2017ApJ...835...51L}.

In the recent paper of \citet{2017ApJ...835...51L} (hereafter Paper~I), an
influence of random variations of the gravitational field of the Galaxy on
apparent celestial positions of extragalactic sources was theoretically
investigated, and statistical characteristics of this process for some
realistic models of our Galaxy were obtained.
It was shown that the jitter, caused by stars or other massive object
moving closely to the line of sight, increases with the observational interval
and reached some maximal value which depends on the sky direction.
In particular, on the scale of 10 years, the deviation of the apparent source position
from the true one can reach several tens of $\mu$as in the direction towards the Galactic Center,
decreasing down to 1-3 $\mu$as at high galactic latitudes.
It is important to note, that in contrast to the random walk of the particle,
the observed fluctuations of the source position occur relative to some ''true'' position.
In general, the functional properties of this effect is determined by
its autocorrelation function and power spectrum (see Paper I for details).

As it follows from calculations, the jitter effect is quite small.
Nevertheless, advances of observational techniques prompt us to pose a
question about its observability. It can be important  for at least  two reasons.
Firstly, the detection of such a jitter poses a fundamental limit on the
astrometric accuracy.
Secondly, a comparison of jitter parameters with
theoretical predictions will help us in the future to validate
the model of the Galaxy used for computations.

A number of environmental factors -- such as mismodelling of path delay in
the neutral atmosphere, mismodelled crustal deformations caused by the
mass loading \citep{r:aplo}, imperfections of the Earth rotation model
\citep{r:erm}  -- affects the accuracy of the absolute radio
astrometry. Although some authors claimed that the accuracy of the
VLBI absolute astrometry can reach 0.05~mas \citep{r:icrf2}, we adhere a more
conservative estimate of the accuracy floor at the level of 0.15~mas.
If not a single object but a pair of objects is observed, the environmental factors are diluted roughly
proportional to the objects angular separation, and thus the accuracy of
the differential astrometry can reach dozens of $\mu$as per single epoch
\citep{r:reho14,r:mvi10}. It should be noted that the differential
astrometry cannot provide positions of observed objects, only a difference in
positions.

This work is a second one in the cycle of papers dedicated to the study of
the gravitational noise of the Galaxy and its influence on the positional
accuracy of extragalactic sources. Here, we consider a possibility to detect
this jitter effect using high-precision radio interferometric observations.
The problem to be solved is formulated in Section~\ref{sec:problem}. In
Section~\ref{sec:error budget}, we estimate the uncertainties of the expected effect
due to an imperfect knowledge of parameters of the Galaxy models, the mass and
velocity distribution functions of deflecting bodies, as well as due to an algorithm of the numerical calculations.
The other possible effects that can cause the observed offset between the two sources
are briefly discussed in Section~\ref{sec:other noises}.
Section~\ref{sec:simulation} presents the main steps in the experiment
simulation. Obtained results and some observational issues are discussed
in Section~\ref{sec:summary}.

\section{The problem formulation}
\label{sec:problem}

Let's consider two groups of bright extragalactic sources -- ``target'' and
``control'' samples. Each sample includes $N$ closely spaced pairs of
sources. Sources of the target sample are located at $|b|\le1.5^{\circ}$ and
$|l|\le20^{\circ}$ (where $l$ and $b$ are galactic longitude and latitude, correspondingly),
i.e.  within the Galactic plane
close to the  Galactic Center, where the expected value of the standard deviation of the jitter effect
(hereafter ``jitter std'') is maximal.
Sources of the control sample are located at high galactic latitudes,
$|b| \ge 30^{\circ}$,
where the jitter std is  predicted to be significantly lower. Measuring
arc lengths between pairs for these two samples and performing a statistical
analysis of the data obtained, we expect to detect a systematic increase of
the  standard deviation in arc lengths of pairs in the ``target'' group with
respect to the ``control'' one on a time scale of several years.

For illustration purposes, we show one particular realisation of simulated samples of extragalactic sources
on the map of the conditional standard deviation of the angular jitter
for the observational interval of 10 years (Fig.\,\ref{Figure.1}).
From this figure it is easy to see that in the absence of any additional noise,
besides the jitter effect, the standard deviations (and dispersions) for
these samples should  differ by several times. Thus, in this
``ideal'' case, it is possible to establish the difference between two samples
of sources at the high significance level. However, the jitter effect will be
observed against a background of various noises, both instrumental and
astrophysical, that can prevent its detection in real data.

Moreover, before simulating an experiment, we need to assess
uncertainties associated with the theoretical calculations of the jitter std provided in Paper~I.
The latter are  based on the present-day mass function (PDMF) of stars, the
velocity and spatial distribution of these stars (the Galaxy model) as well
as some simplifying assumptions. It is obvious that uncertainties of
parameters of models and functions will affect the amplitude of the effect.
Thus, our primary task is to calculate the whole budget of possible errors
arising due to these uncertainties, as well as due to the algorithms of
numerical calculations. Below, we consider all these issues which affect detectability of the jitter effect.

\section{Uncertainties of the theoretical modelling}
\label{sec:error budget}

According to Paper~I, the spatial, velocity and PDFM
distribution functions are independent, therefore the uncertainty for the
jitter std can be estimated as summation in quadrature of uncertainties of
each of the distribution functions. Below we consider them separately.

\subsection{Galactic models}
\label{sec:galactic models}

Our non-accurate knowledge of the  structure of the Galaxy is the first
and the main problem which leads to the modelling uncertainties. In Paper~I, all
calculations were carried out for two models of the density
distribution of the matter in the Galaxy: 1) the ``classical''
Bahcall-Soneira model \citep{1980ApJS...44...73B,1986ARA&A..24..577B} and 2)
the more  realistic model of the Galaxy of \citet[][(their model
2)]{1998MNRAS.294..429D}. For convenience, let's call the latter one as the
`basic' model. It is necessary to note that this model is only one out of
four models which were obtained by these authors from the analysis of the
same observational data. Therefore, in order to assess how much the choice of
the Galaxy model affects our estimates of the jitter std across the sky,
we performed its calculations for three
other models from \citet[][models 1, 3, and 4]{1998MNRAS.294..429D}, and
for the Galaxy model from  \citet{2017MNRAS.465...76M}, which was constructed
using latest observational data.

Using the same technique as in Paper~I, we constructed a set of maps of the
jitter std for different models of the Galaxy.
A comparison of these maps shows that, depending on the used model,
variations of the jitter std are about several percent at high galactic
latitudes, increasing  to $20-25$\% in the central part of the Galaxy and low
latitudes. Such an increase in uncertainty is connected with difficulties of the
parametrization of the central part of the disk and the bulge components. For
the following estimations, we used 25\% as a conservative value of the
uncertainty of the jitter std due to our non-accurate knowledge of
the exact structure of the Galaxy.

\subsection{Present-day mass functions}
\label{sec:the present-day mass function}

It is well known that the mass function of Galactic stars can not be
determined directly from observations. Observable quantities, e.g., the
luminosity function or the surface brightness, are transformed into the mass
function through the mass-age-luminosity relation. Thus the mass function is
obtained within the framework of a given theory of a stellar evolution.

In our calculations, we used universally recognized present-day mass
functions for the disc, halo and bulge stars from papers of
\citet{1997A&A...328...83C, 2003PASP..115..763C}. Expressions for PDMFs of
different galactic components include two or three parameters, which are
determined within  given uncertainties. Obviously, the scatter in the
PDMFs parameters affects the results obtained
in Paper~I. We varied the parameters of PDMFs randomly
within the intervals of their uncertainties and estimated the resulting uncertainty on the jitter std to be 9-15\%,
depending on the sky direction and contribution of
different galactic components. For our
following calculations, we use 15\% as a conservative value of the
uncertainty on the jitter std due to our non-accurate knowledge of the mass functions.

\subsection{Stellar velocity distribution}
\label{sec:star's velocity distribution}

A velocity distribution of stars, used in Paper~I,  depends on two
parameters: the escape velocity from the Galaxy and the dispersion of the
stellar velocities, which are different for different galactic components.
Variations of the escape velocity within 10\% of its value 500\,km\,s$^{-1}$
do not have any noticeable impact on the jitter std. Variations of
the stellar velocity dispersions within their uncertainties, reported by
\citet{2016ARA&A..54..529B} for different galactic components, lead to the
variations in the jitter std which are below 5\%.

Thus the uncertainty due to our non-accurate knowledge
of the stellar velocity distribution can be conservatively estimated as 5\%.

\bigskip

Finally, uncertainties of the conditional
standard deviations or the autocorrelation functions can be as large as
20\%, depending on the choice of the minimal impact parameter of a deflecting
body in respect to the observable extragalactic source (see Paper~I for
details).

Summarizing all above, an overall `theoretical' uncertainty on the jitter std,
arising due to different types of the modelling uncertainties, can be conservatively
estimated as 35\%.

\section{Other noises}
\label{sec:other noises}

Let's consider  other possible effects that affect measurements of the arc length
between two sources.
They are as follows:
1)~thermal noise; 2)~the contribution of path delay in the ionosphere;
3)~the contribution of path delay in the neutral atmosphere.
Atmospheric errors grow approximately
linearly with the source separation, therefore, the shorter
the separation, the better. \citet{r:mvi10} provided realistic estimates of
differential astrometry errors as a function of source separation.
At the $2^\circ$ separation, the accuracy of 30~$\mu$as can be achieved
with the interferometer baselines of $\sim$2000--3000 km (Honma, private communication).
\citet{r:reho14} noted that random errors of differential astrometry for
the $1^\circ$ source separation can even reach  $\sim$10 $\mu$as
for the interferometer baseline of $\sim$8000 km, although systematic
errors are usually higher.

One has to take into consideration several other factors. First, the
structure for many AGNs is changing with time, that is related to
their flaring activity \citep[see, e.g.,][and references
therein]{r:mojave17}. During a flare, a component is ejected from the core
regions, travels with the relativistic speed, then fades out and disappears.
Presence of an extended jet, if left unaccounted for, typically contributes to estimates of source positions at the level
of 30--100~$\mu$as. However, if a jet has a compact component,
in extreme cases its contribution to the source position may surpass 1~mas according to
recent results of \citet{r:gaia3}. Changes in the source structure due
to the evolution of jet components, if unaccounted for, result in
the change of source position estimates. This effect can  be modelled using
source maps, however the question on the residual errors of the source
contribution accounted that way remains open.

Another phenomenon is a core-shift. The core position is shifted along the jet
due to  self-absorption, and this shift is frequency dependent. The
effect was predicted by \citet{r:bk79} theoretically and then confirmed from
observations \citep[see, e.g.,][and references therein]{r:kov08,r:sok11}. A
typical value of the core-shift at 8~GHz is about 200 $\mu$as.
Recently \citet{r:pla19a} has demonstrated a variability of the
core-shift related to the flaring activity. The core-shift is reduced at
high frequencies, although as \citet{r:cs15_43} has shown, it is still at the
level of 100 $\mu$as at 15~GHz. Thus, multi-frequency observations are
needed  to evaluate the core-shift and its evolution.

Another non-stability of the source position and broadening of the image
can be caused by  scattering in the interstellar medium
\citep{r:pus15,r:laz08}. The broadening is common at low galactic
latitudes, and in the extreme cases a source cannot be detected at long baselines
at 1--8~GHz due to broadening. Since the astrometric accuracy of VLBI is
reciprocal to the baseline length, a loss of long baselines reduces
astrometric accuracy. Moreover, clouds of the interstellar medium
can change the broadening, and such variations associated with extreme
scattering events may happen at scales of months
\citep{r:pus13,r:cimo02,r:fiedler87,r:fiedler94}. The broadening is
reciprocal to frequency squared, and observations at high frequencies,
22~GHz and higher, substantially mitigate this effect.

The impact of these effects can be minimized, if one observes close pairs of sources
(separated by no more than 1--$2^\circ$) at high frequencies, i.e. at
22~GHz or higher, and performs simultaneous observations at several frequencies (at least, at two)
to evaluate the core-shift and estimate the remaining frequency-dependent
ionospheric contribution.

\section{Experiment simulation}
\label{sec:simulation}

In this section, we carry out an experiment simulation
guided by the above recommendations on the observation strategy.

\begin{figure*}
\centering
 \includegraphics[width=\textwidth]{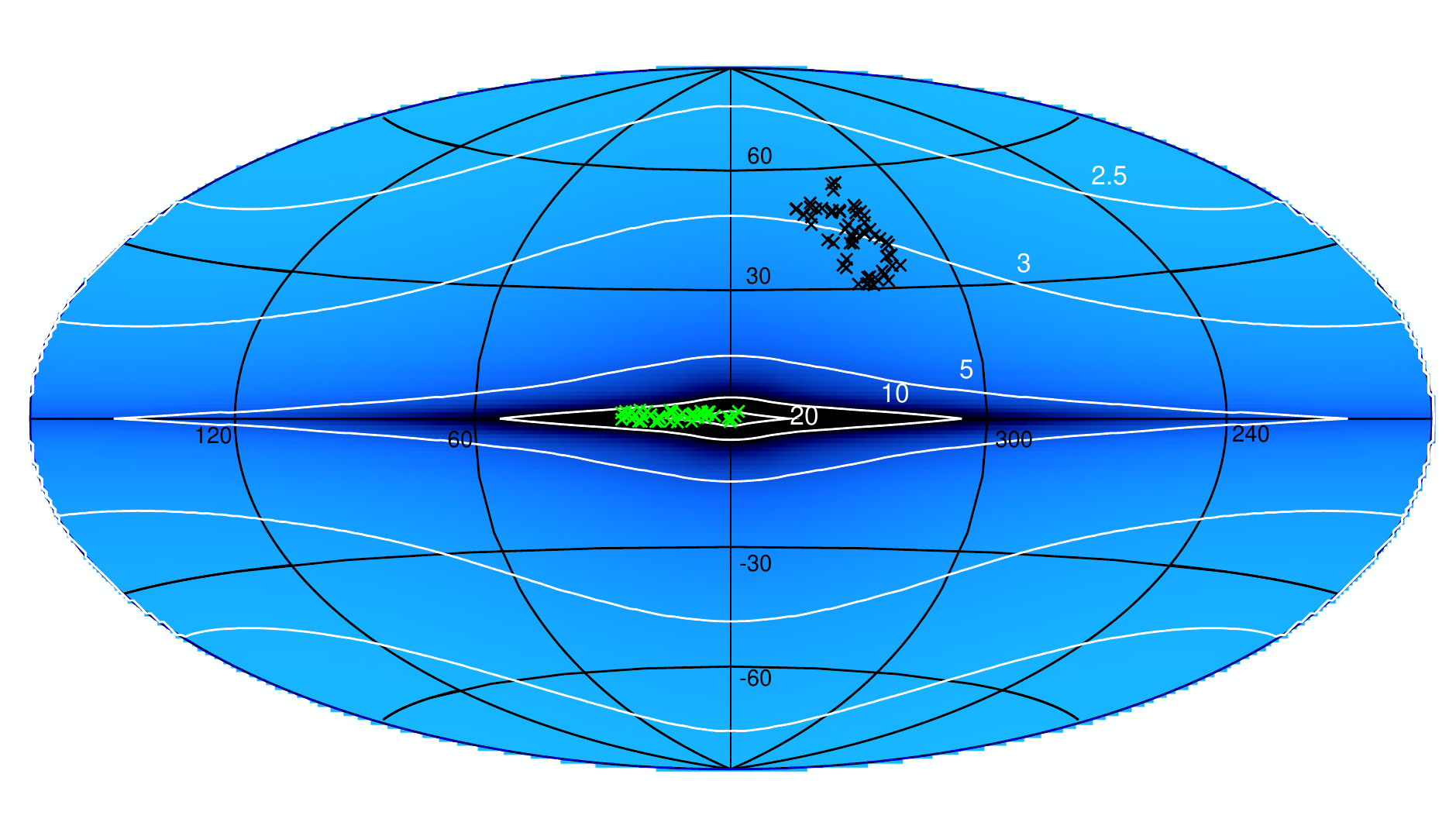}
\caption{Map of the conditional standard deviation of the angular jitter (in
$\mu$as) for the observational interval of 10 years. White lines show contours
of the jitter angle $\alpha =$ 20, 10, 5 and 3 $\mu$as. Positions of the
target (in green) and control (in black) sources are marked as crosses.
\label{Figure.1}}
\end{figure*}

We start  with generating two samples of distant
sources. The target sample, consisting of $N$ source pairs, is created at random
in the central sky region with coordinates  $|b|\le1.5^{\circ}$ and
$|l|\le20^{\circ}$. The separation in the pair varies from
$1^{\circ}$ to $2^{\circ}$. A lower limit of separation is set due to absence
of  spatial correlation of the jitter effect. As discussed in previous Section, the accuracy of the
differential VLBI astrometry starts to degrade noticeably when the
separation exceeds $2^\circ$ due to short-term variations in path delay
through the atmosphere. The initially specified arc lengths in the pair of
sources is considered as the ``true''
one\footnote{Of course, in real observations this parameter is unknown.
Here it is needed only as a starting point for further modelling.}.

The choice of the longitude/latitude for the control sources is determined by the minimal value of the effect.
In general, control sample sources should have the latitude greater than $30^{\circ}$ (see Fig.\,\ref{Figure.1}).
Here, the control sample is created in the sky region
$-70^{\circ}\le l \le -20^{\circ}$ and $30^{\circ} \le b \le 60^{\circ}$
in the same way as the target sample.

As soon as two samples are generated, at the next stage we need to simulate
``observational'' data, taking into account the studied jitter effect as well as
an influence of different types of  noise. The experiment
simulation includes three steps:
1)  first, we generate an
array of ``measured'' arc lengths in each pair in the absence of any noise
other than the jitter effect;
2) then, we synthesize
three types of noise - white, flicker and red - that represent different types of observational and instrumental noises;
3) finally, we produce and analyze a noisy signal.

Now we consider all these steps in details.

\subsection{Signal generation}
\label{sec:signal generation}
The main purpose of this subsection is to simulate the jitter effect
with the predicted statistical parameters in the absence of any other noise.

Let's assume that we observe $N$ closely spaced pairs of sources of the target
and control samples $K$ times during $T$ years. After $T$ years, we expect to have $N$
time sequences consisting of $K$ measured arc lengths between sources of the
$i$-th pair $l_{i}(t_{j})$, where $j = 1, ..., K$,  in a given
(target or control) sample. To simulate the noise-free observation of the
jitter effect, we generate arrays of ``measured'' arc lengths $l_{i}(t_{j})$
under the following assumptions:

\begin{enumerate}

\item the jitter std of sources in each
$i$-th pair, $\alpha_{i1}$ and $\alpha_{i2}$, are independent
(since the minimum separation between sources in a pair is chosen to avoid a spatial correlation);

\item  values of $\alpha^2_{i1}$ and $\alpha^2_{i2}$ for a given time
interval between observations are calculated according to the predicted
autocorrelation function of the jitter effect from Paper~I;

\item the generated arc lengths $l_{i}$ between sources in the $i$-th pair are
drawn  from the Gaussian distribution with an average value equal to the
``true'' arc length $l^{true}_{i}$ (since the mathematical expectation of the jitter vector is
zero), and with the std  $\alpha_{tot}(t)$, which depends on time;

\item in its turn, $\alpha_{tot}(t)$ is also distributed over the Gaussian
with the mean of  $\mu = \sqrt{\alpha^2_{i1}(t)+\alpha^2_{i2}(t)}$ and
the width of $\varepsilon \simeq$ 0.35$\mu$, where 35\% is the `theoretical' uncertainty on the jitter std discussed in Section\,\ref{sec:error budget}.

\end{enumerate}


\begin{figure*}
\centering
 \begin{tikzpicture}
 \node[above right] (img) at (0,0) {
 	\includegraphics[width=0.19\textwidth]{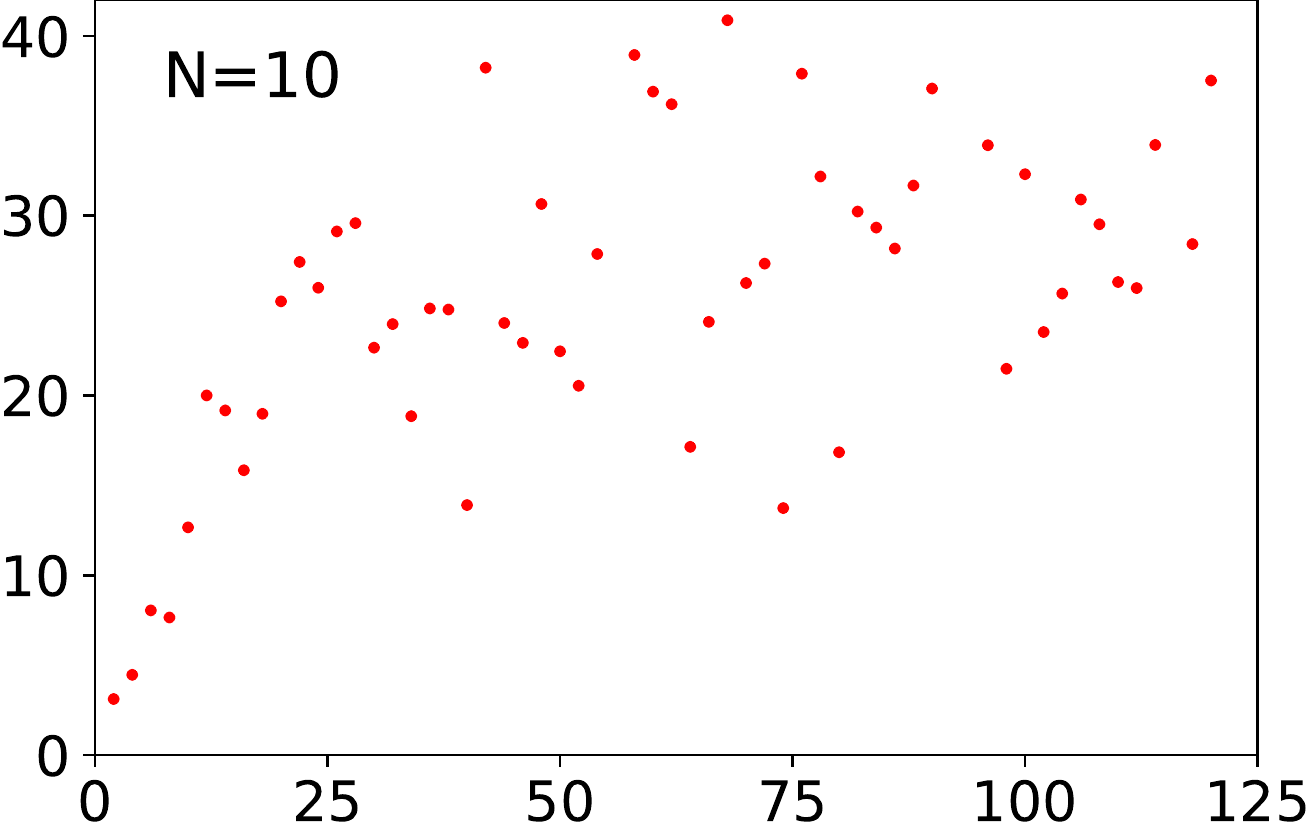}
 	\includegraphics[width=0.19\textwidth]{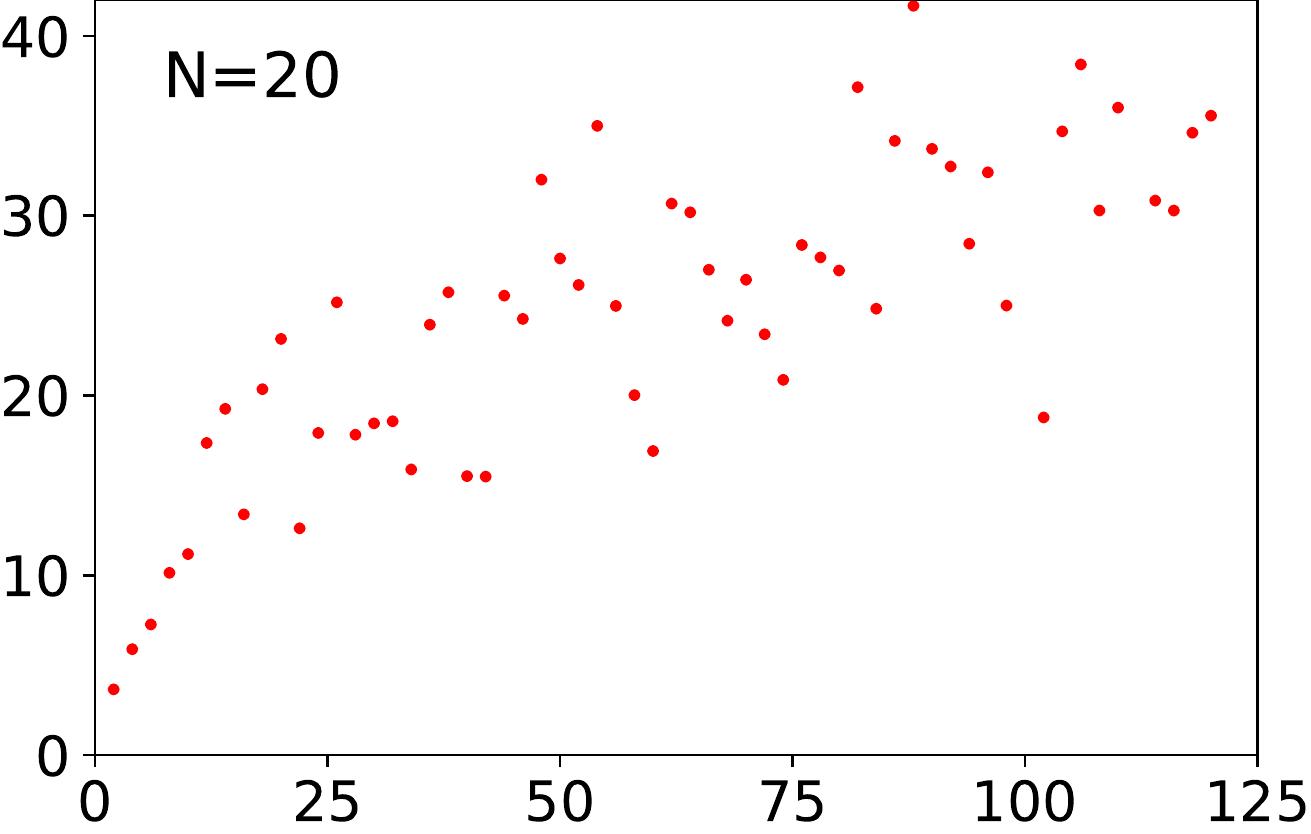}
 	\includegraphics[width=0.19\textwidth]{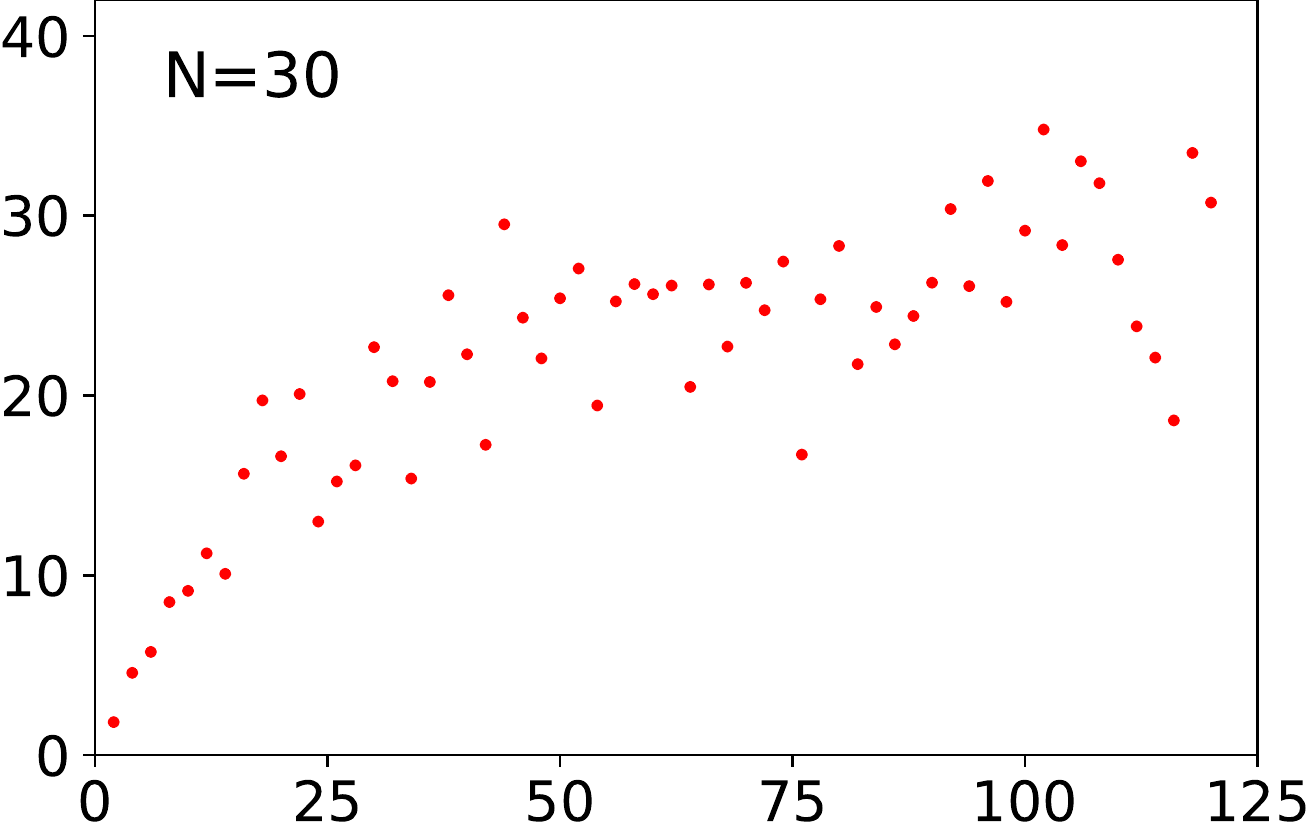}
    \includegraphics[width=0.19\textwidth]{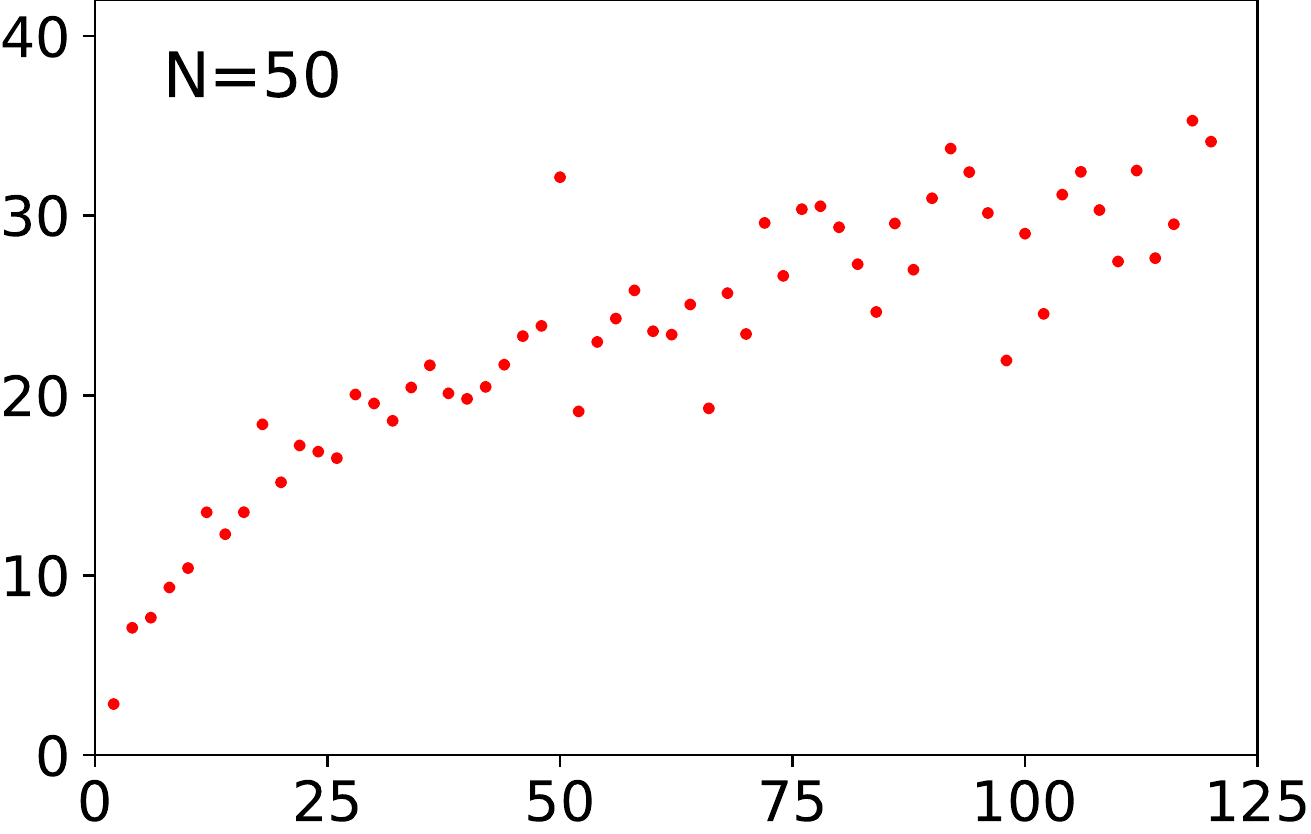}
    \includegraphics[width=0.19\textwidth]{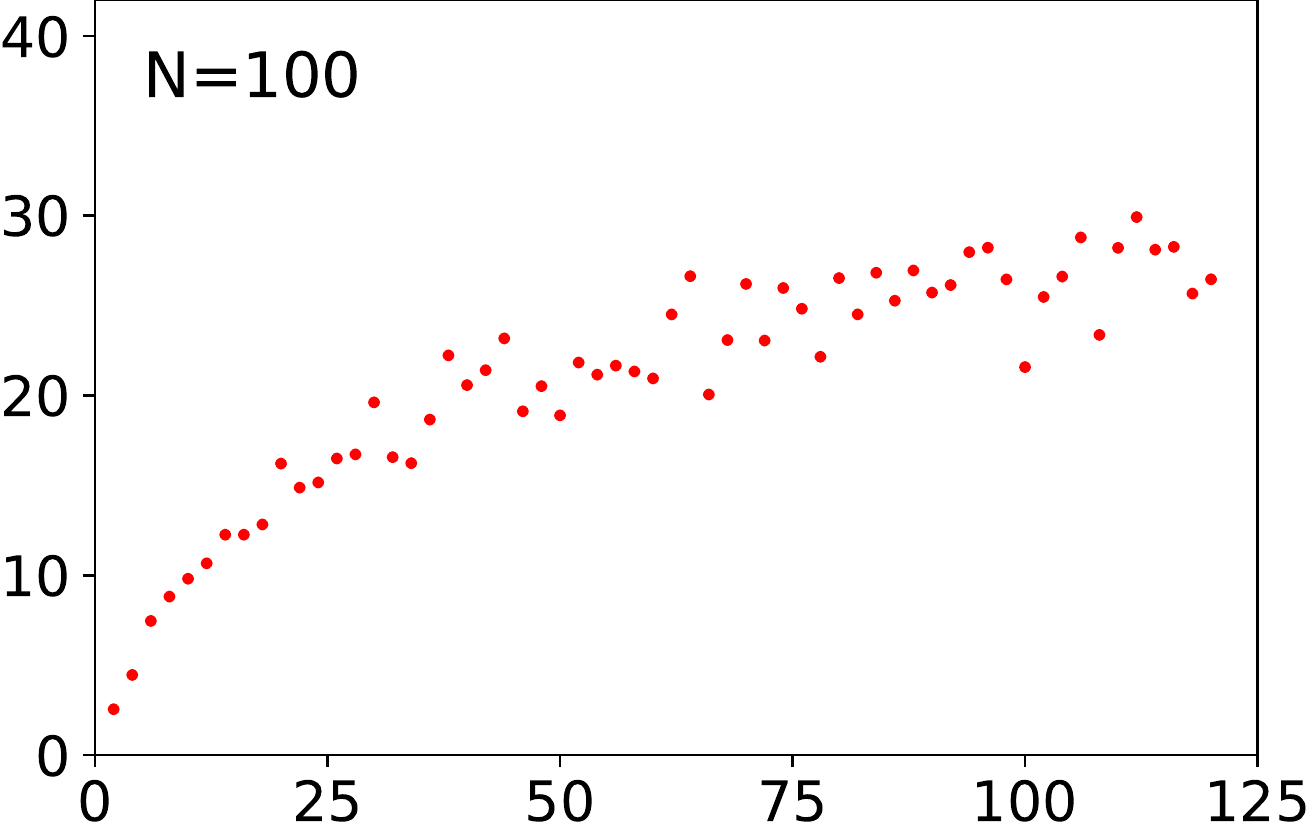}
 };	
 \node[label={[rotate=90]right:jitter std, $\mu$as}] at (-0.4,0.1) {};
 \end{tikzpicture}
 \centering
 \begin{tikzpicture}
 \node[above right] (img) at (0,0) {
 	\includegraphics[width=0.19\textwidth]{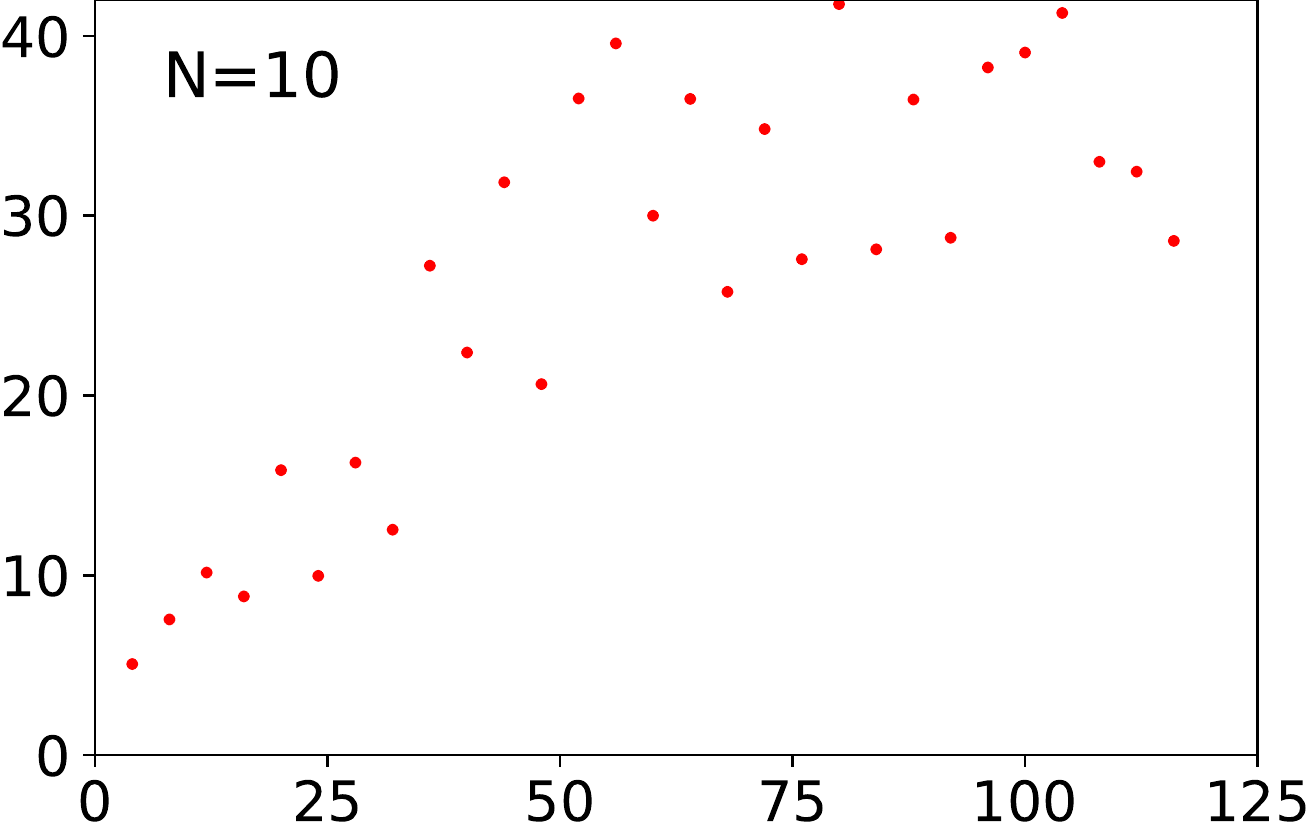}
 	\includegraphics[width=0.19\textwidth]{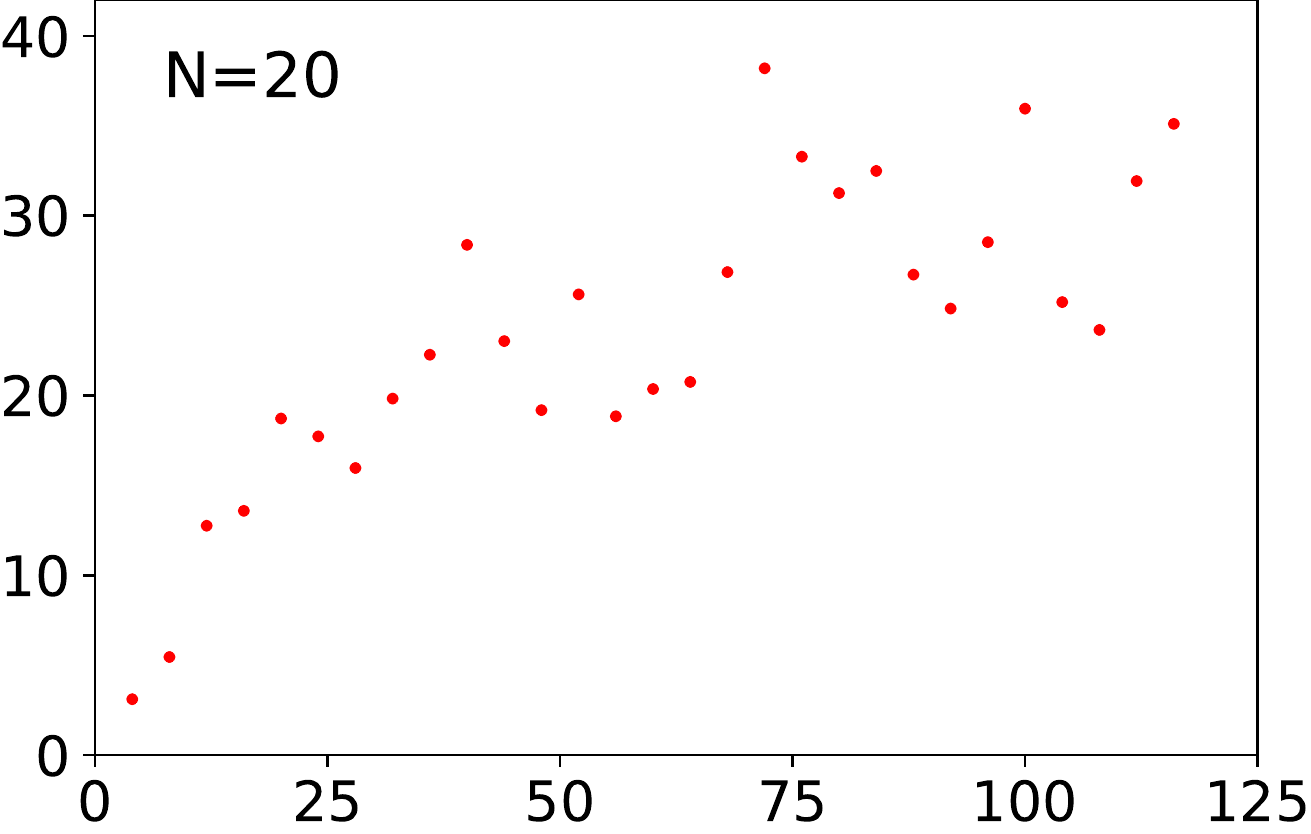}
 	\includegraphics[width=0.19\textwidth]{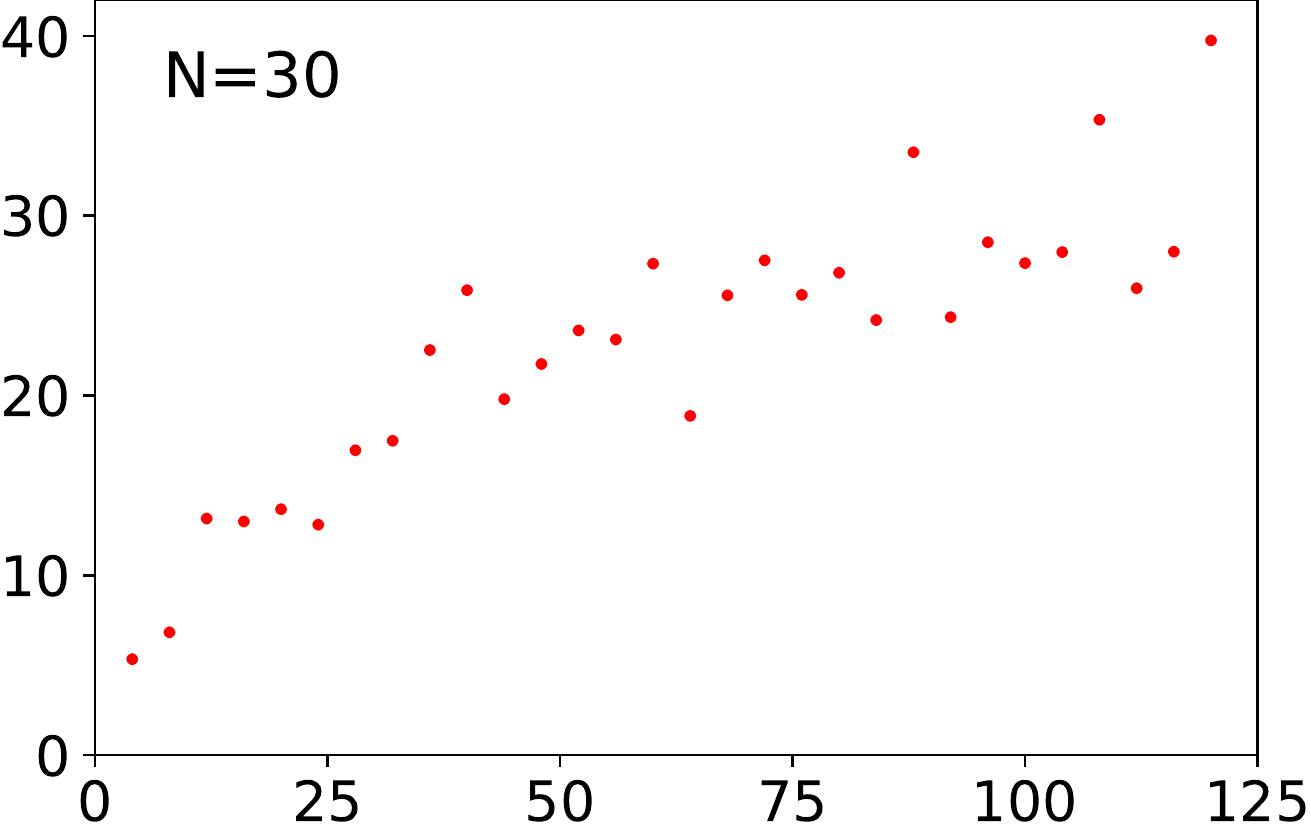}
    \includegraphics[width=0.19\textwidth]{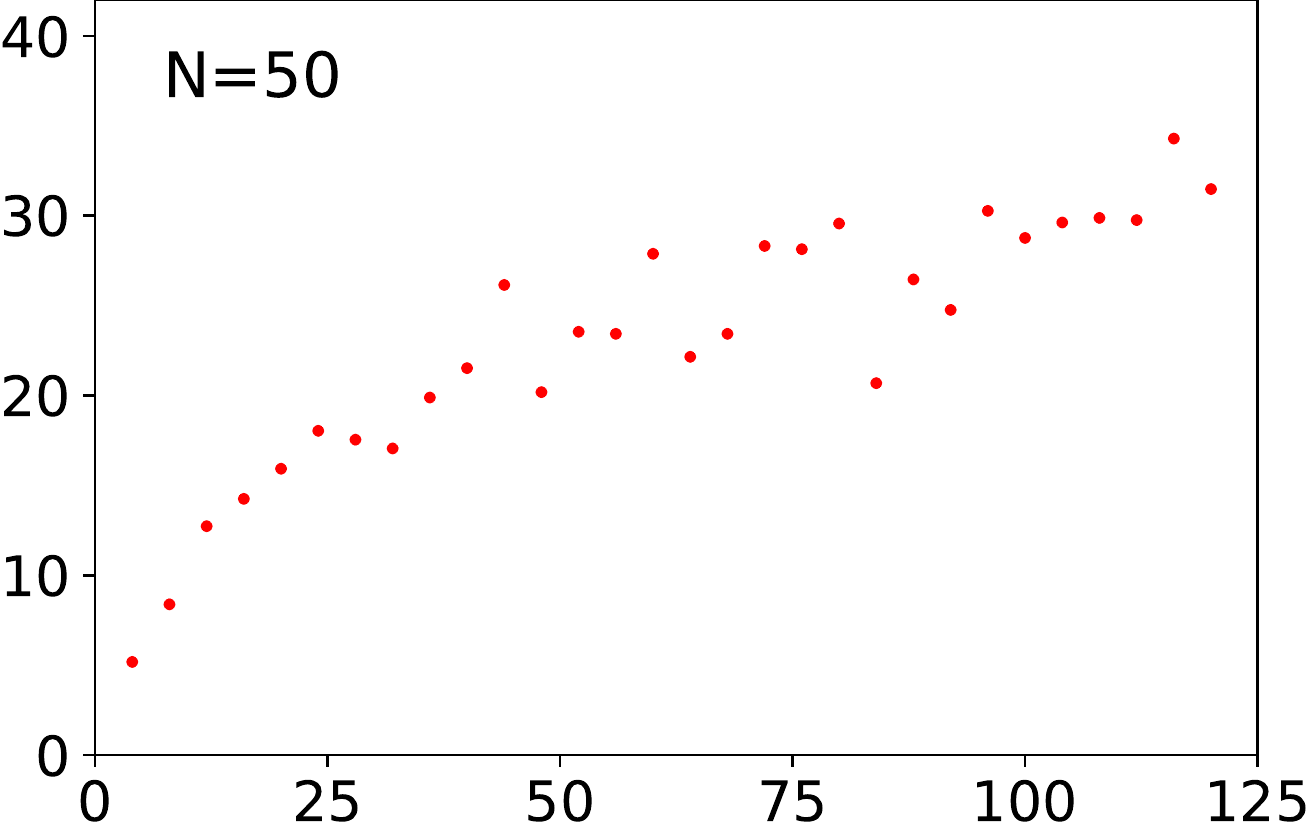}
    \includegraphics[width=0.19\textwidth]{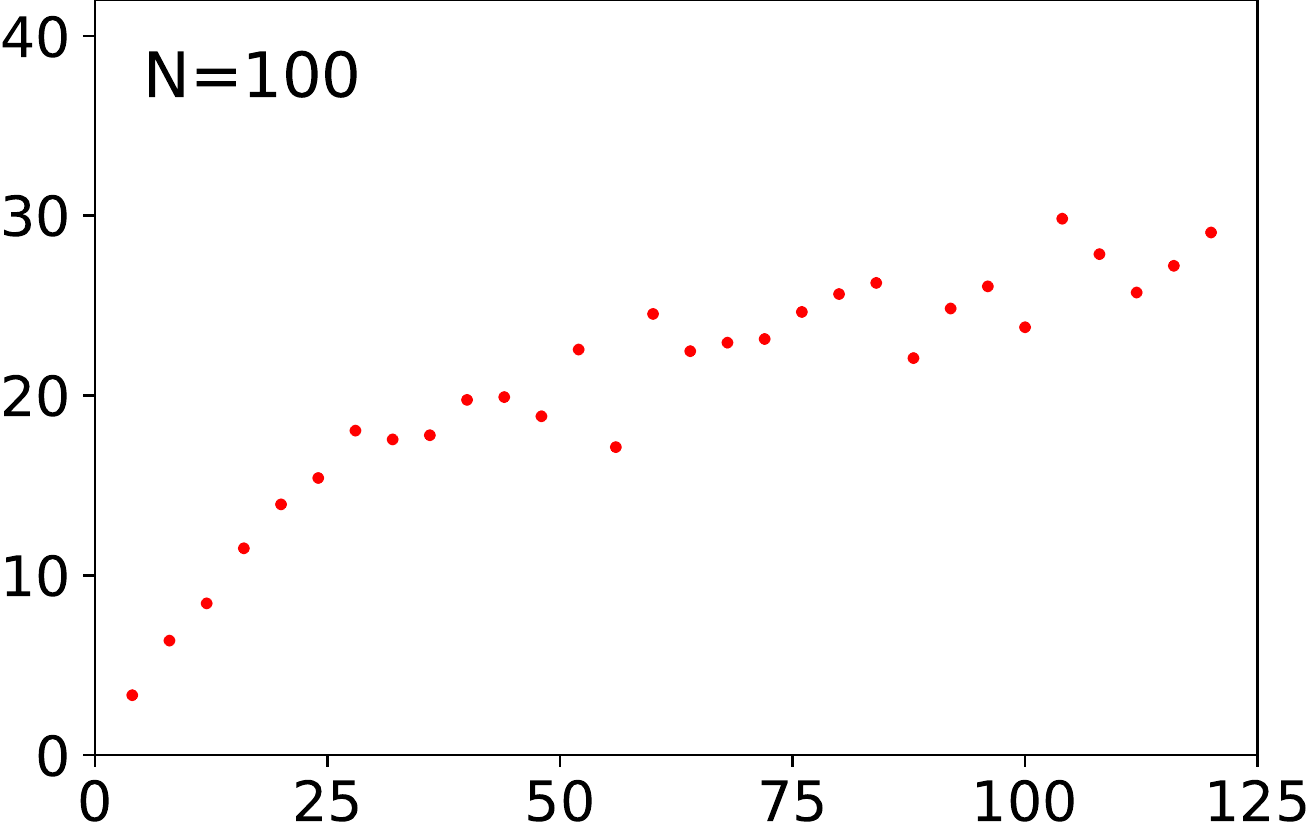}	
 };
 \node[label={[rotate=90]right:jitter std, $\mu$as}] at (-0.4,0.1) {};
 \node[rotate=-90,label=months] at (7.7,-0.1) {};
 \end{tikzpicture}
\caption{The jitter std as a function of time (in months) for different
number of source pairs (N=10, 20, 30, 50 and 100) from the target sample.
Only one random realization is shown. In the upper row, the sampling interval
between observation is equal to 2 months, in the lower row, the interval is equal to 4 months.}
\label{fig:Nsources}
\end{figure*}


As soon as we have the array of ``measured'' arc lengths $l_{i}(t_{j})$ for
the $i-$th pair of sources, we can calculate the average arc length for a given
pair as  $ <l_{i}> = \frac {1}{K} \sum_{j=1}^K l_{i}(t_{j})$
and subtract this value from each ``measured''  arc length. Thus, for each
$i$-th pair of sources we obtain a sequence of arcs $\Delta l_{i}(t_{1} ),
..., \Delta l_{i}(t_{K} )$, where $\Delta l_{i}(t_{j} )= l_{i}(t_{j} )- <l_{i}>$.
This sequence is called a signal for the $i$-th pair. A similar procedure is
performed for all the $N$ pairs from the target sample. As a result, we obtained the
matrix [$K$x$N$] of arcs $\Delta l_{i}(t_{j})$, which is one of the
possible realizations of our measurements for the target sample.
This matrix was used to calculate the standard deviation for this random realization in
the appropriate way\footnote{Throughout the paper, we calculate the standard
deviation as  $\sigma = \displaystyle \sqrt{ \frac{\sum_{k=1}^{K} (x_k -
<x>)^2}{K-1}}$, where $\{x_1,x_2,...,x_K\}$ are the observed values, $<x>$ is
the mean value, and $K$ is the number of observations in the sample. The
variance is denoted as  $\sigma^2$.}.

It is obvious that for each of the possible realizations
the calculated standard deviation of the arc lengths will be different.
To define the mean value of the jitter std and its scatter,
we performed 100 such realizations.
We tested that increasing the number of realizations
above 100 doesn't lead to any change in derived quantities at 1 $\mu$as level.
So, the spread of the average value of the obtained jitter std determines
the range of expected values of that can be obtained
in the experiment with no extra-noises.

The described above algorithm is carried out for the pairs of the control
sample as well. As a result, we obtain 100 matrices [$K$x$N$] of
arc lengths $\Delta l_{i}(t_{j} )$  and calculate the average value of the
jitter std and its spread for both target and control samples.

\begin{figure*}
	\centering
\begin{tikzpicture}
\node[above right] (img) at (0,0) {
 	\includegraphics[width=0.31\textwidth]{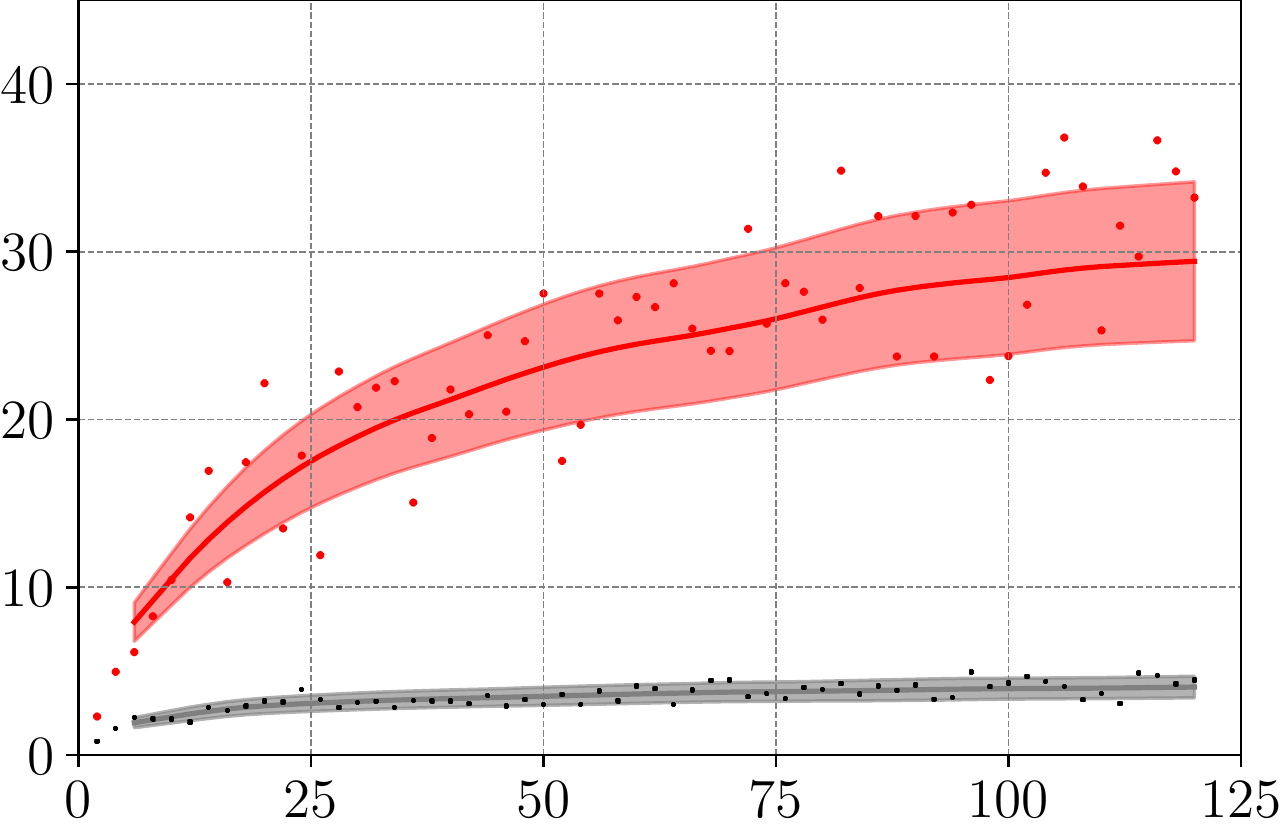}
 	\includegraphics[width=0.31\textwidth]{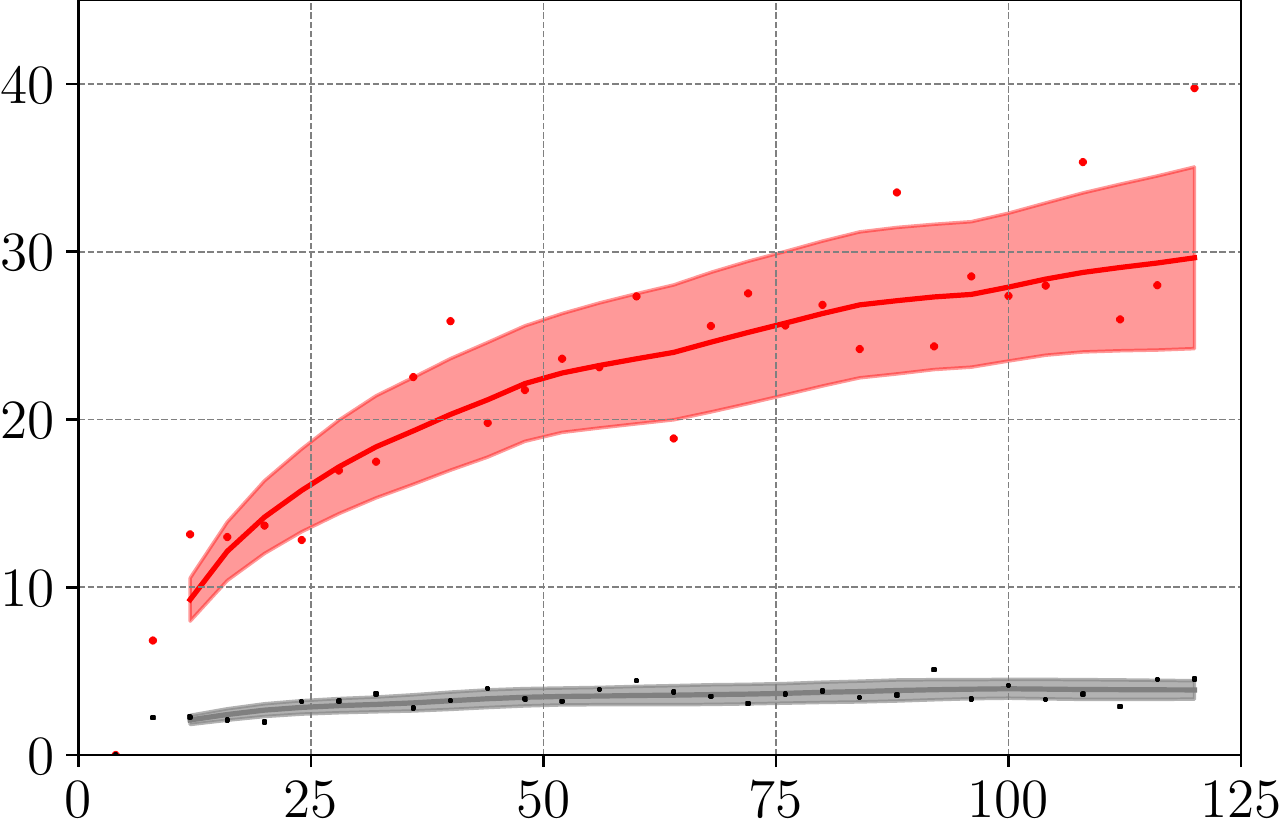}
 	\includegraphics[width=0.31\textwidth]{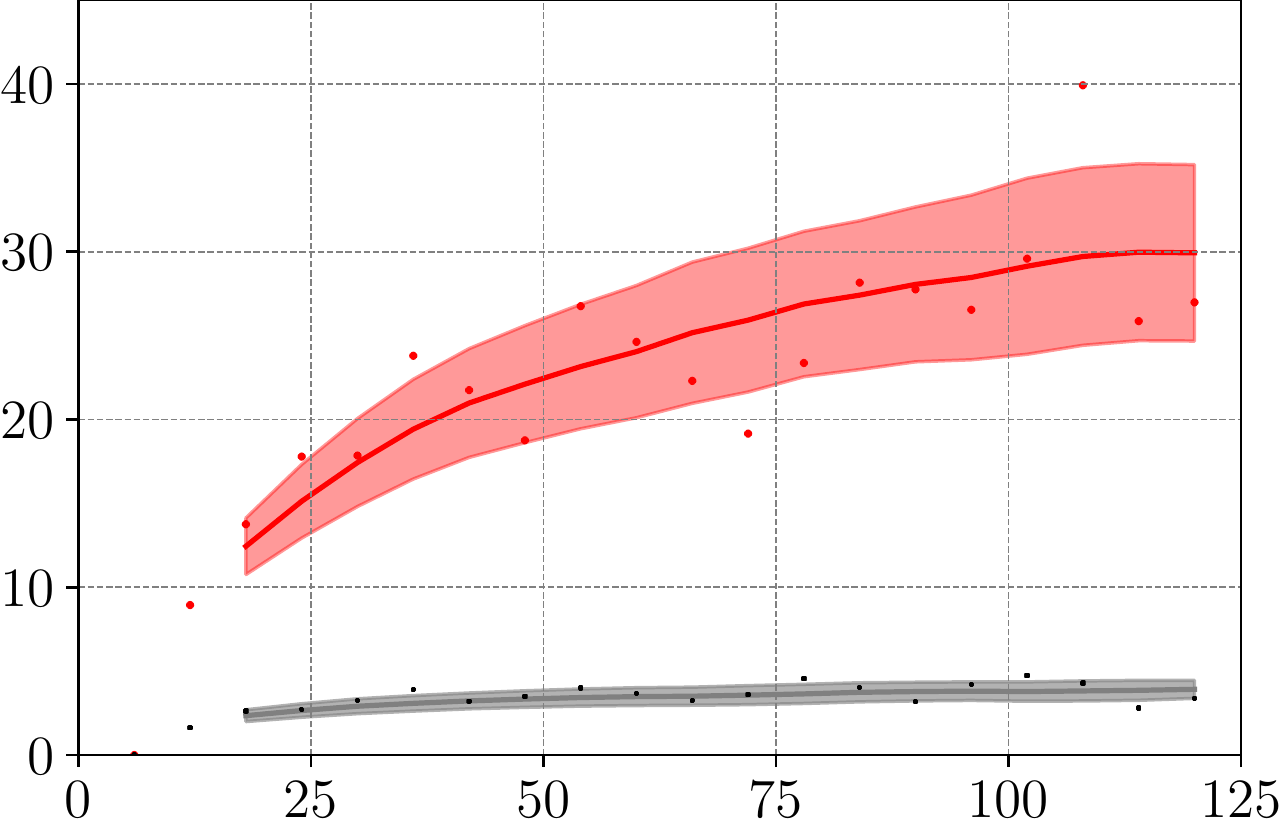}
};	
\node[label={[rotate=90]right:jitter std, $\mu$as}] at (-0.4,1) {};
\end{tikzpicture}
\centering    	
\begin{tikzpicture}
\node[above right] (img) at (0,0) {
 	\includegraphics[width=0.31\textwidth]{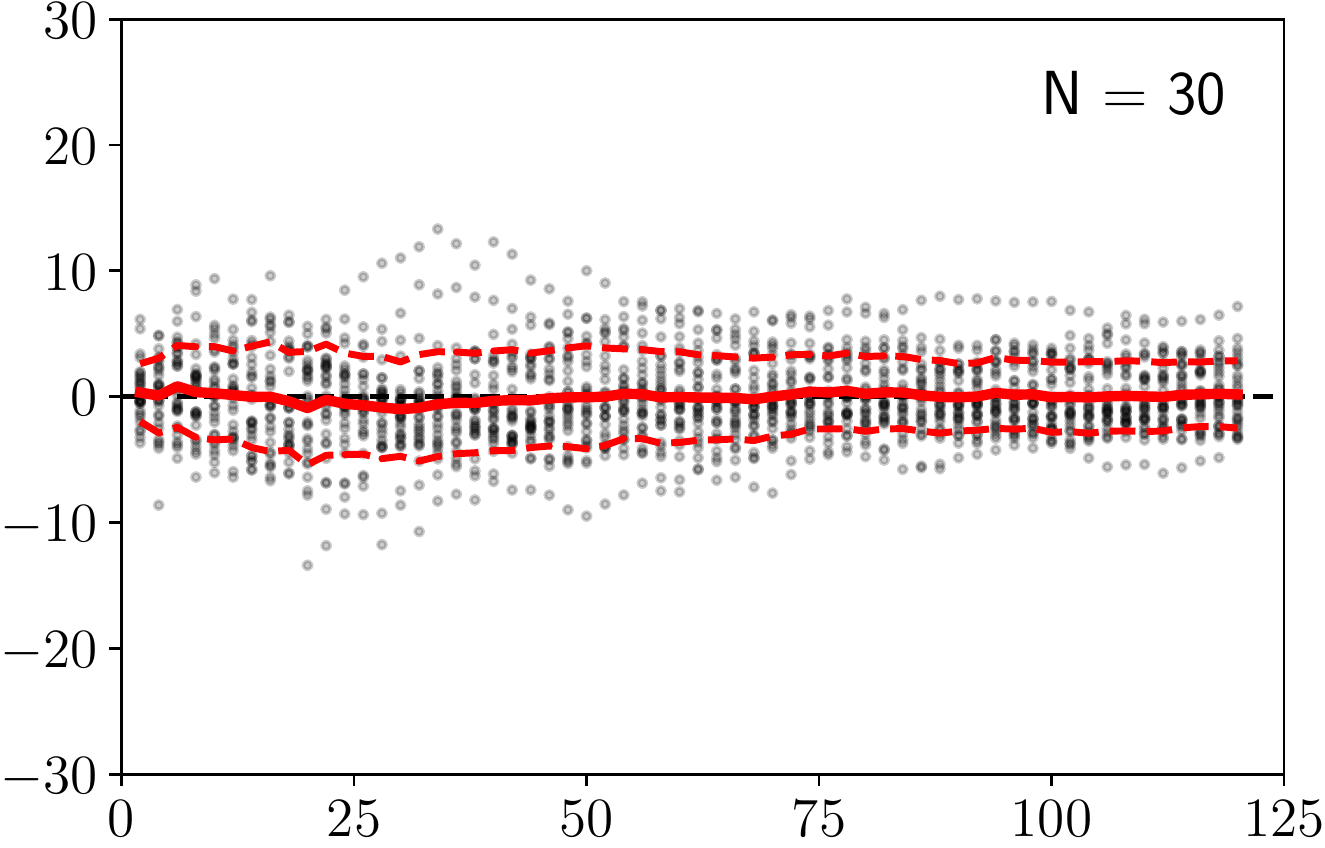}
    \includegraphics[width=0.31\textwidth]{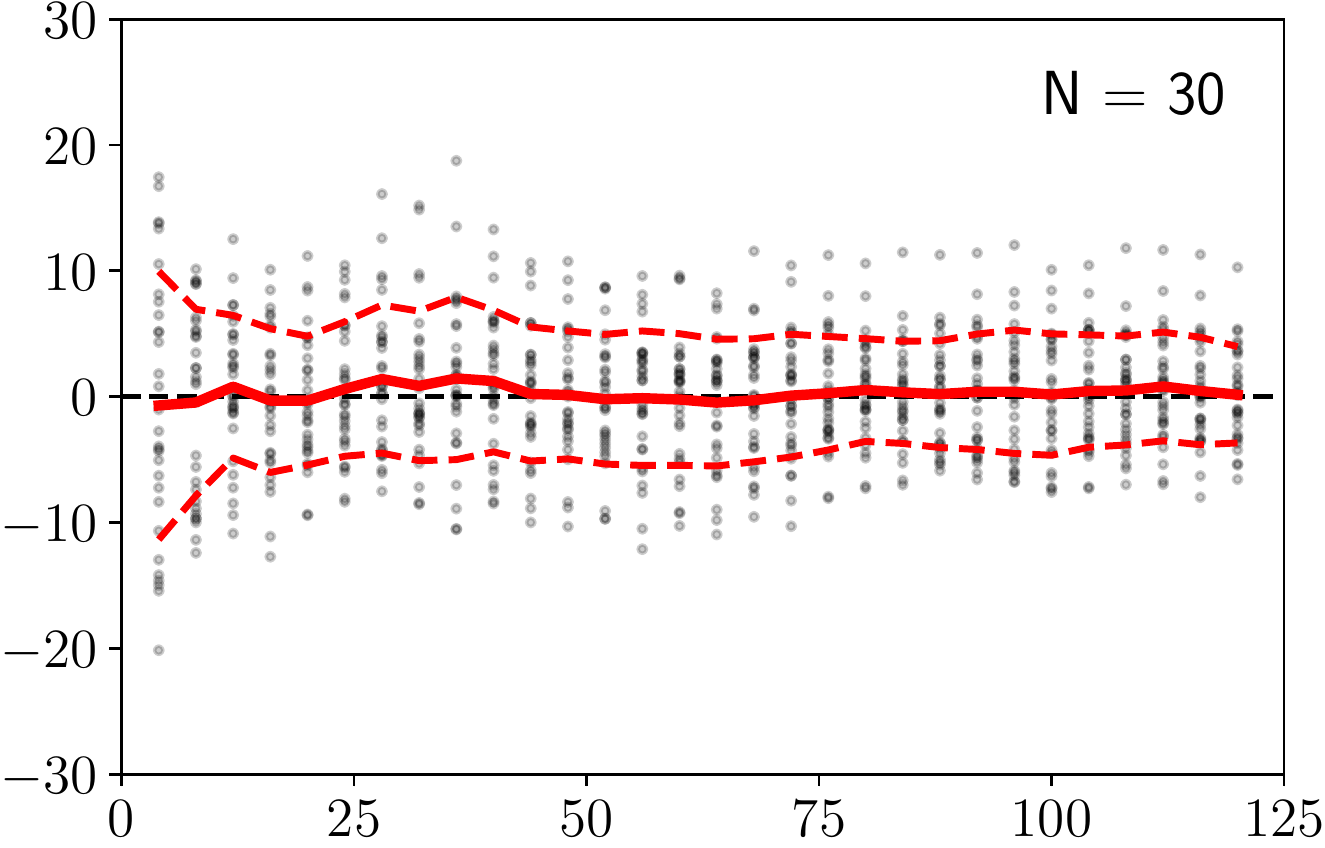}
    \includegraphics[width=0.31\textwidth]{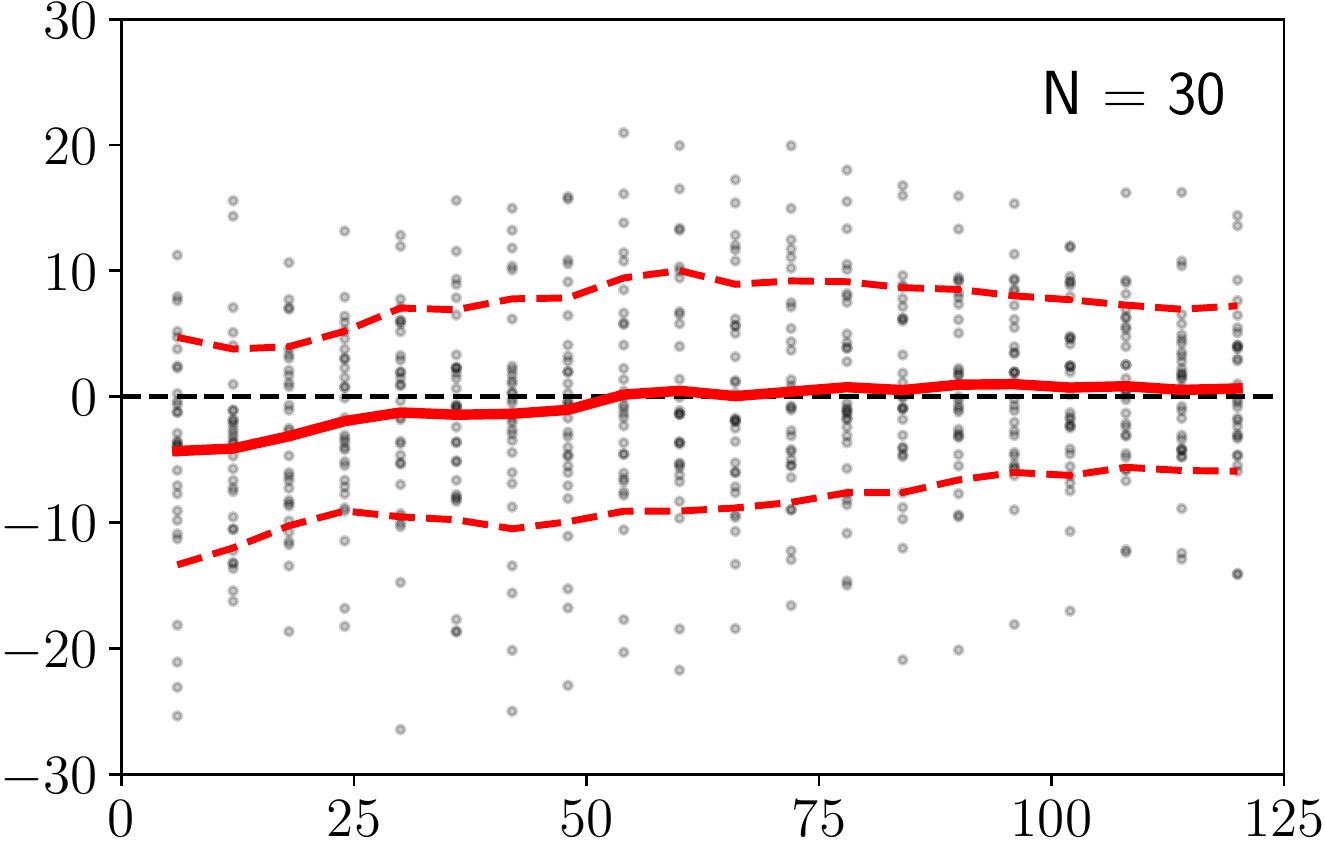}
};
\node[label={[rotate=90]right: $\Delta = <l> - l^{true}$, $\mu$as}] at (-0.4,0.15) {};
\node[rotate=-90,label=months] at (7.7,-0.1) {};
\end{tikzpicture}	
\caption{\textit{The upper panels:} the jitter std as a function of time. From left to right: observations are made each 2, 4, and 6 months, respectively. Red and grey thick lines show the mean value of the jitter std for 100 realizations of target and control samples, respectively.
The uncertainties are shown by the red and grey shaded areas, for the target and control sources. Red and grey dots represent one random realization for the target and control sources, respectively. \textit{The lower panels:} the cumulative moving average arc lengths between 2 sources converge with time to its true value. Grey dots show measurements for each of 30 pairs of sources. Red lines show the mean value of $\Delta = <l> - l^{true}$ and the uncertainty (by dashed lines) on this mean (see text for details).}
\label{fig:Figure.2}
\end{figure*}

Now it is important to estimate how many pairs of sources, what duration and the duty cycle
of observations are needed for the adequate statistical analysis.
The scatter of the average value of the jitter std obviously depends on the number of pairs of sources as $1/\sqrt{N}$.
Even in the idealized noise-free experiment, for 10 pairs of sources,
an expected growth with time of the standard deviations of measured arc lengths
could be buried under random fluctuations (see Fig.~\ref{fig:Nsources}),
while for $N=30$, the  trend with time can be clearly recognized.
There is no doubt that the larger the number of source pairs, the more accurate the statistics.
Nevertheless, for observational reasons, it is necessary to determine a sufficient number of pairs for the purposes of our experiment (Fig.~\ref{fig:Nsources}).
So hereafter, we concentrate on the target and the control samples,
consisting of $N=30$ pairs of sources each.
Figure~\ref{fig:Figure.2} represents the mean value of the jitter std
with no extra-noises and its scatter for 30 pairs of target (red lines) and control (grey lines)
sources as a function of time for different sampling intervals between observations: 2, 4 and 6 months.
For the target sample, the mean value of std increases with time and reaches $\simeq (25\pm4$) $\mu$as
after 5 years of observations and $\simeq (30\pm5$) $\mu$as after 10 years.
For the control sample, the mean value of std varies slightly with time
and approximately equals to ($3\pm1$) $\mu$as after 10 years of observations.
The slope of the signal power spectrum equals to $\simeq -2 $ as derived in Paper~I.

Note, that the sampling interval does not noticeably affect the jitter curve (top panel of Figure~\ref{fig:Figure.2}),
since the smooth curves were obtained by averaging over a large number of realizations of the experiment.
However, in a real experiment, only a single sequence of observations (as opposed to 100 realizations in our calculations) is available. In such a case, increasing the frequency of observations obviously leads to more accurate estimates of mean pairwise arc lengths. Figure~\ref{fig:Figure.2} (lower panels) illustrates how for one random realization the cumulative moving average arc length between sources in a pair converges with time. Measurements for each of 30 pairs of sources are marked with grey dots. Red lines show the mean value of $\Delta = <l> - l^{true}$,
where $<l>$ is the cumulative moving average arc lengths at the time instance $t$, and the uncertainty (one standard deviation) on this mean.

\subsection{Noisy signal}
\label{sec:noise generation}

In the previous section, we have simulated an ideal experiment with no
extra-noises except for the noise from the jitter effect, which is considered to be the useful
signal here. Now we consider the more realistic situation when this useful signal is
``spoiled'' by the noise.

Different effects that cause the non-stability of the source
position and proposed a strategy to minimise them were discussed in
Section\,~\ref{sec:other noises}. Not all the arising noises/effects are well
understood and can be completely removed from observational data. Here, we
assume that the jitter is spoiled with some kind of noise  left after the
data cleaning process. Since the jitter effect is different in nature from
other astrophysical noises (such as a flaring activity of AGN, core shift,
scattering in the interstellar medium, etc), as well as from instrumental
noises, we do not expect any correlation between the useful signal (the
jitter effect) and the noise. Thus, we assume that the noise is additive. In
the following analysis we consider three types of noise: white noise, flicker
noise (the spectrum with $1/f$) and red noise (the spectrum with $1/f^2$) as
the most common  in the analysis of time series in Astronomy and Meteorology
\cite{1978ComAp...7..103P,1981ApJS...45....1S,2013arXiv1309.6435V}.

\begin{figure}
\begin{center}
 \includegraphics[width=\columnwidth]{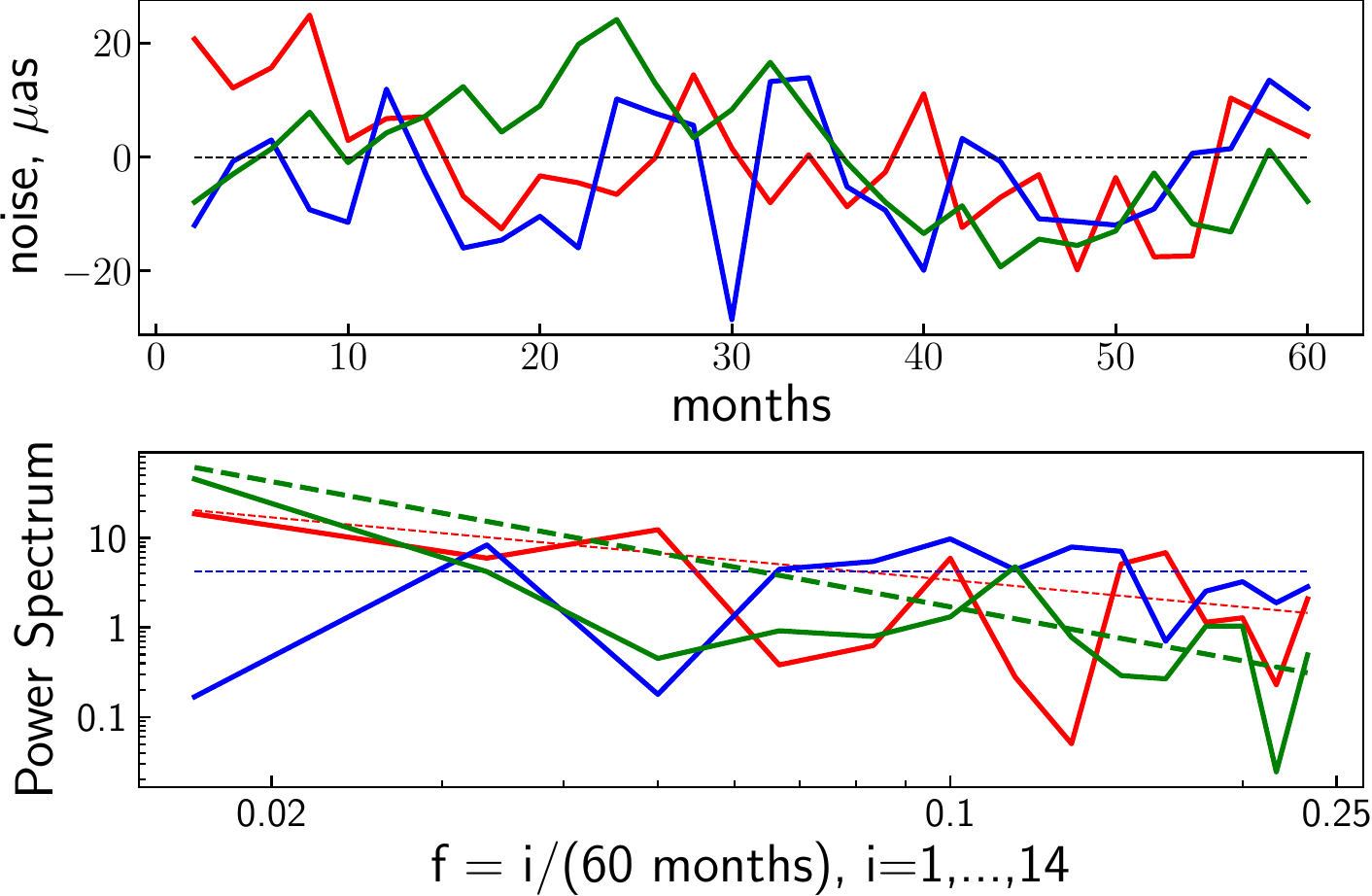}
\caption{Illustration of different types of noise considered in the paper.
The upper panel shows the dependence of the noise amplitude in $\mu$as versus time.
White noise is shown in blue, the flicker noise is in red, the red noise is in green.
The bottom panel shows the power spectrum. The blue dashed line is a constant (white noise),
red is $1/f$ (the flicker-noise), green is $1/f^2$ (the red noise), where $f = i/(60$ months),
$i=1,...,14$ is a (discrete) Fourier transform frequency.
\label{Figure.3}}
\end{center}
\end{figure}

The white noise time sequence
for each pair of sources is drawn from a Gaussian distribution
with a zero mean and a given variance $\sigma_{n}^2$. A colored
noise (with a power law power spectrum) is generated using the algorithm
described  in \cite{1995A&A...300..707T}. The average value of the
noise amplitude in each time sequence is set to zero, the mean variance of
noise time sequences is $\sigma_{n}^2$. As an example, one realization of
different types of the noise with some given value of $\sigma_{n}^2$ is shown in Fig.~\ref{Figure.3}.
The upper panel of the figure shows the dependence of the noise amplitude on time,
where the white noise is shown in blue, the flicker noise is in red and the red
noise is in green. The bottom panel of the figure shows the power spectrum,
where the blue dashed line is a constant (the white noise), red
corresponds to the spectrum with $1/f$  (the flicker-noise), green
corresponds to the spectrum with $1/f^2$ (the red noise).

As in the case of the signal generation, 100 realizations of the noise
amplitude matrix for three types of noise with a given power spectrum and
variance were obtained for simulating the ``spoiled'' signal.

To obtain a noisy signal, the noise amplitude matrix is added to
the signal amplitude matrix for each realization.
As a result, we have 100 matrices [$K$x$N$] of ``noisy'' arc lengths
$\Delta l_{i}(t_{j} )$ for both target and control samples, where $N$=30.
These matrices were analyzed in the same way as
for the matrices of pure signal (see Section \ref{sec:signal generation}).


\begin{figure*}
 \centering
 \begin{tikzpicture}
 \node[above right] (img) at (0,0) {
 	\includegraphics[width=0.45\textwidth]{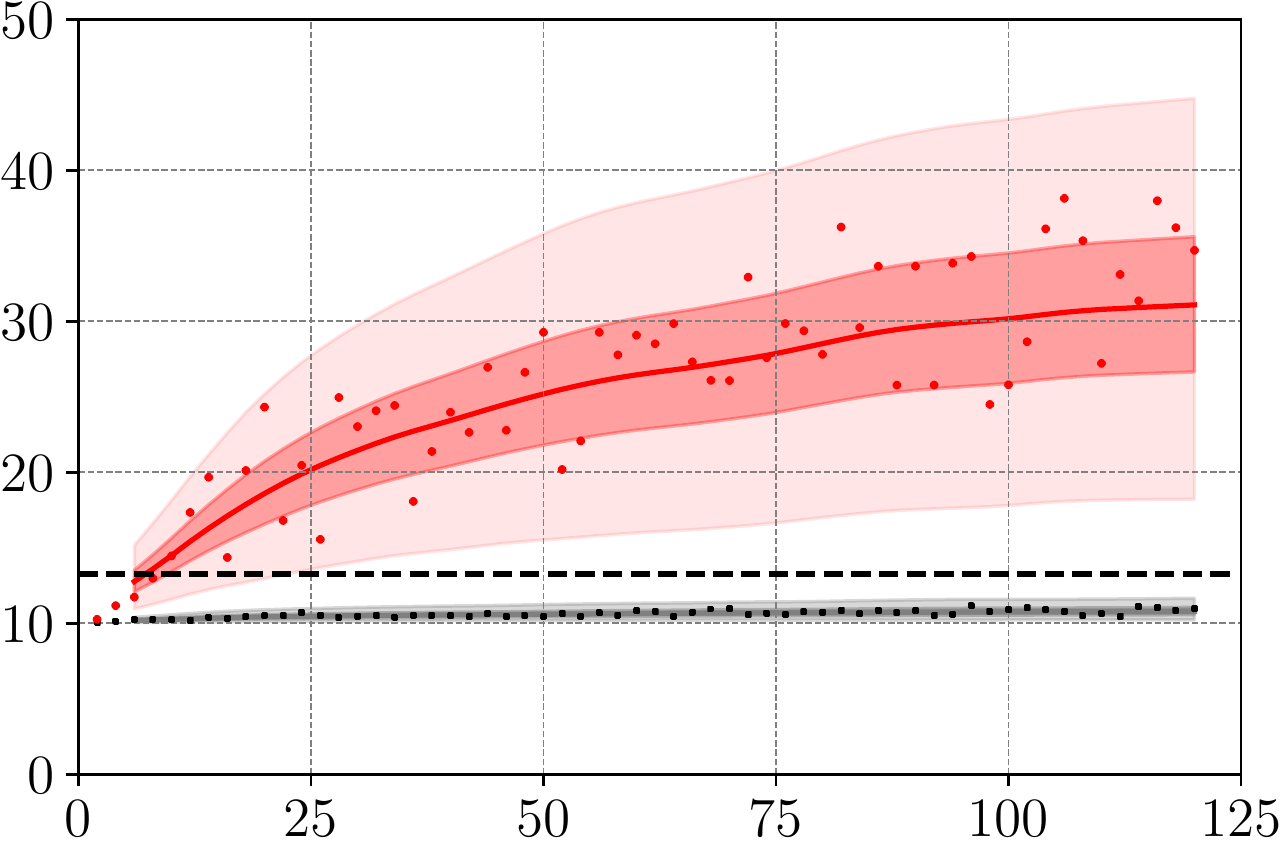}
 	\includegraphics[width=0.45\textwidth]{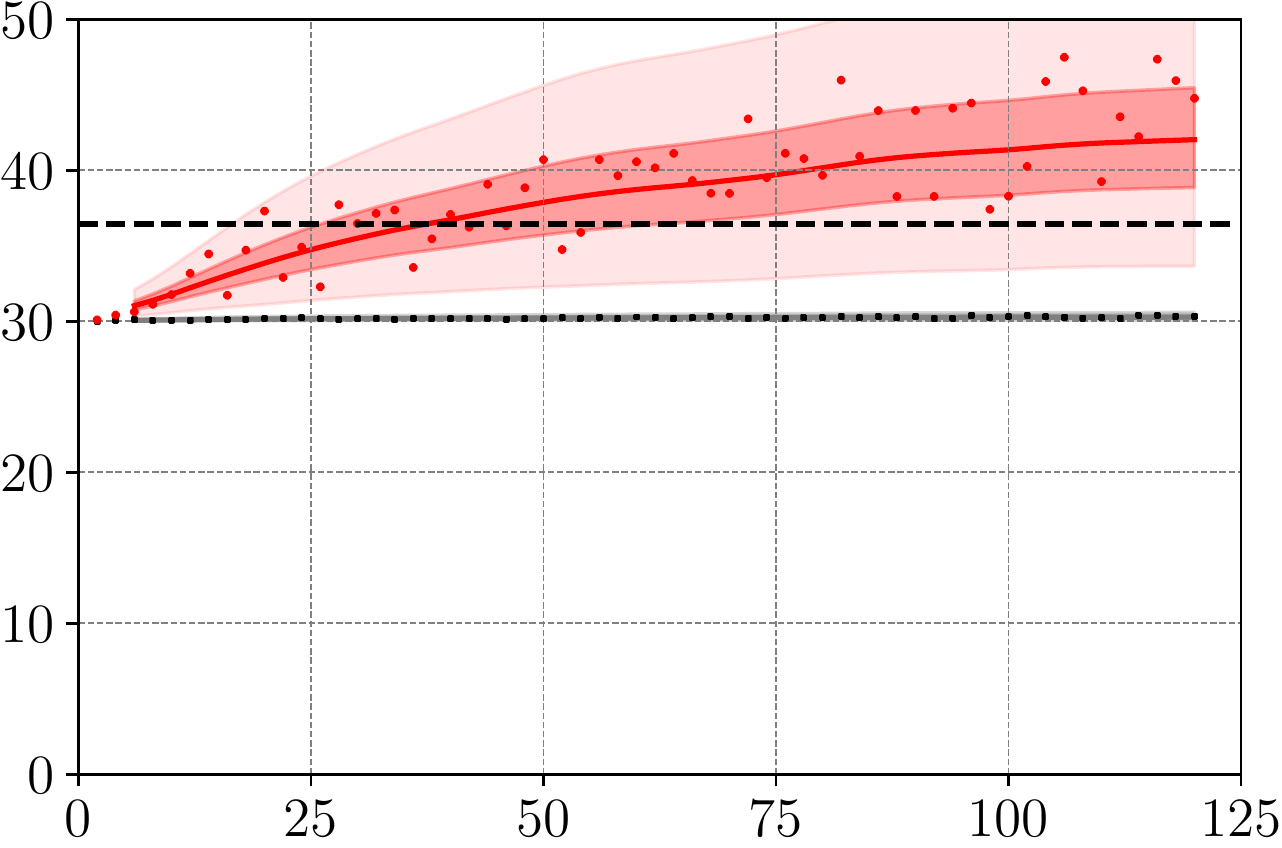}
 };	
 \node[label={[rotate=90]right:jitter std, $\mu$as}] at (-0.4,2) {};
 \node[rotate=-90,label=months] at (3.5,-0.1) {};
 \node[rotate=-90,label=months] at (11.7,-0.1) {};
 \end{tikzpicture}
\caption{Prospects for detecting the jitter effect in `noisy' observations.
The thick lines show how the mean std of measured arc lengths between sources
in pairs increases with time.  Shaded dark areas represent the uncertainty on
this mean (i.e. one standard deviation), shaded light areas represent 3
standard deviations. Observations are made each 2 months. Left and right
panels show the curves for the jitter signal spoiled by an additive noise
with the mean variance of $\sqrt{\sigma^2_n}$ = 10 $\mu$as and 30 $\mu$as,
correspondingly. The statistical properties were calculated for a sample of
100 realizations. One random realization is shown with dots. Red and grey
color are used for the target and the control sample, correspondingly.
According to the Fisher criterion, the target and control samples can be
distinguished at the 3$\sigma$-level, if the light red shaded area lies above
the dashed thick horizontal line.
\label{Figure.5}}
\end{figure*}


We assume that the signal is spoiled by only one type of noise, either white
or colored, with $\sigma_{n}$ = 10 $\mu$as and $\sigma_{n}$ = 30 $\mu$as. As
mentioned in Section\,\ref{sec:other noises} these values roughly
correspond to the level of differential astrometry errors for different
interferometer baselines \citep{r:reho14}. Note that all types of the noise
give the similar result as long as: (i) calculations are performed for the
fixed number of source pairs, and (ii) the noise variance remains the same
for different colors of the noise. Taking this into account we show in
Fig.~\ref{Figure.5} the std and its spread assuming that the signal is
spoiled by the white noise only. We conduct averaging over 100 realizations
of a noisy signal, as in the analysis of the signal with no extra-noises.
Thick lines show how the mean std  of measured arc lengths between sources in
pairs  increases with time. Shaded dark areas represent the uncertainty on
the mean (i.e. one standard deviation), shaded light areas represent three
standard deviations. Statistical properties were calculated for a sample of
100 realizations. Red and grey dots are the std of the noisy signals for one
random realization of the target and control samples, respectively. According
to the  Fisher criterion, one can distinguish between the target and control
samples at least at the  3$\sigma$-level, if light red shaded area lies above
the horizontal dashed line.

It can be seen that for our choice of $\sigma_{n}$, the standard deviation and its spread
for the control sample are entirely determined by the noise.
Figure~\ref{Figure.5} also shows that for the target sample the scatter
of the standard deviation increases with the duration of experiment
which is owing to the presence of the ``jitter'' effect, in contrast to the control sample,
where the time variation is practically constant with time.
As follows from our calculations, on the scale of $\sim$2 years, it would be possible to detect a systematic increase
in the std of measured arc lengths of pairs of target sources compared to the control ones
at the $3\sigma$-level (according to the Fisher criterion) if the accuracy of differential astrometric observations is 10 $\mu$as.

\section{Summary and discussion}
\label{sec:summary}

Motions of stars and compact objects in our Galaxy cause local fluctuations
of its gravitational field which lead to jittering of apparent celestial positions of distant sources.
In Paper~I, it was found that the jitter amplitude depends on a direction in the sky
and can reach several tens of microarcsecond.

In this paper, we  considered the possibility of detecting the jitter effect on the basis
of the  theoretical predictions of its value  and
the current accuracy of the differential astrometry.
We simulated long-term measurements of the arc lengths of the closely spaced source pairs divided into two groups:
``target'' and ``control''. Target sources lie in the direction to the central part of the Galaxy,
where the expected jitter effect is maximal,
while control sources are located at high galactic latitudes,
where the predicted jitter amplitude is minimal. Different types of physical and
instrumental noises were taken into account in the form of the additive extra-noise
(white, red, and flicker) with a constant dispersion in time.

It was shown that on the scale of $\sim$2 years, it  is possible to detect  a systematic increase
in the std of measured arc lengths of pairs of target sources compared to the control ones
at the $3\sigma$-level (according to the Fisher criterion) if the noise
dispersion $\sigma_{n}$ is 10~$\mu$as. If $\sigma_n$ is 30~$\mu$as,
then the target and control samples will differ only at the 2$\sigma$-level on the scale of 10 years,
that can be considered as a possible hint for the jitter effect.
These values of $\sigma_{n}$ roughly correspond to the level of differential
astrometry errors for different interferometer's baseline \citep{r:reho14}.
Note the accuracy of $\sim 20~\mu$as is necessary to achieve the $3\sigma$-level on 10 years interval.

We make a conclusion that measurement of extra jitter is achievable with the
current technology of radioastronomical observations, although a great case
should be taken for assessment of the contribution of systematic errors and
characterization of their power spectrum. The distribution function of the
observed arc lengths has a potential to differ the Galactic models.
As it was shown in Section 3.1, the difference in the jitter values can reach
25\% for the explored models, presumably in central parts of the Galaxy.
Thus, the accuracy of astrometric measurements at the level of
few $\mu$as is required to resolve this task.

We also discussed
other possible effects that can also affect the arc length measurements between two sources.
To make the detection of the astrometric jitter in the Galaxy possible, one should minimize them.
It can be achieved by (1) observing close pairs of
sources (within 1--$2^\circ$ to each other), (2) observing at high frequencies (22~GHz and higher),
and (3) observing at least at two frequencies simultaneously to
evaluate the core-shift and solve for frequency-dependent remaining
ionospheric contribution.
\citet{r:mfo} have shown how this can be done in practice. 
Systems like VERA and KVN that offer simultaneous 22/43~GHz
capabilities and just recently have demonstrated 30~$\mu$as astrometric
accuracy, as well as proposed ngVLA, that will cover the frequency range
from 1.2 to 112~GHz \citep{r:ngvla_ref_des} seem the most promising in these aspects.


\bibliographystyle{aasjournal}

\bibliography{LLL_noise_apj_vP_1}

\end{document}